\documentclass[12pt,oneside,dvipdfm,fleqn]{article}

\usepackage{amsmath}
\usepackage{amssymb}

\usepackage[pdftex]{graphicx}

\usepackage[square,numbers,sort&compress]{natbib}
\usepackage{booktabs}

\begin{document}

\thispagestyle{empty}
\vfill

\begin{center}\textbf{\Large Near-Critical Fluctuations and Cytoskeleton-Assisted Phase Separation Lead to Subdiffusion in Cell Membranes}\end{center}

\vspace{30mm}

\begin{center}
Jens Ehrig, Eugene P. Petrov,\footnote[1]{Corresponding author: petrov@biotec.tu-dresden.de} and Petra Schwille\par
\end{center}

\begin{center}
\textit{Biophysics, BIOTEC, Technische Universit\"{a}t Dresden, Tatzberg
47/49, 01307 Dresden, Germany\\}
\end{center}

\vfill

To appear in \textit{Biophysical Journal} Vol. 99, Issue 12, 2010.

\vfill

\newpage

\section*{Abstract}
We address the relationship between membrane microheterogeneity and anomalous subdiffusion in cell membranes by carrying out Monte Carlo simulations of two-component lipid membranes. We find that near-critical fluctuations in the membrane lead to transient subdiffusion, while membrane--cytoskeleton interaction strongly affects phase separation, enhances subdiffusion, and eventually leads to hop diffusion of lipids. Thus, we present a minimum realistic model for membrane rafts showing the features of both microscopic phase separation and subdiffusion.
\newline\newline

\noindent\textbf{PACS:} 87.14.Cc, 87.15.ak, 87.16.dj, 87.16.dt.
\newline

\noindent\textbf{Keywords:} Lipid membranes, DMPC, DSPC, phase separation, membrane rafts, subdiffusion.

\newpage

\section*{Introduction}

Anomalous subdiffusion in cell membranes is an intriguing phenomenon whose molecular origins are still a subject of debate \cite{Saxton, Destainville, Saxton_response}. From the general viewpoint, the phenomenon is related to (dynamic) microheterogeneities in the properties of the cell membrane creating a rugged energy landscape for the diffusing protein or lipid molecule. The current understanding of membrane microdomains embraces the concepts of lipid rafts \cite{Simons} and cytoskeleton-based picket fence \cite{Fujiwara}. The general concept of membrane rafts \cite{Pike} implies a dynamic interplay between the membrane local composition and phase, and local membrane-protein interactions, which can result in local lipid demixing and phase separation. This view is supported by the recent experimental evidence demonstrating that local lipid demixing and phase separation can be induced by crosslinking \cite{Hammond}, change in the local membrane curvature \cite{Roux}, and interaction with a cytoskeleton \cite{Liu}. Additionally, the presence of the membrane-associated actin network leads to a shift in the membrane phase transition temperature and, potentially, to broadening of the phase transition \cite{Liu}.

The efficiency of external perturbations in changing the local properties of the membrane should be strongly enhanced in the vicinity of the membrane critical point. This is indeed supported by recent experimental observations \cite{Sorre}. Moreover, it was suggested that dynamic microdomains in the cell membrane are nothing but near-critical fluctuations in the local composition and phase of the membrane  \cite{Veatch}. All this is in agreement with the observation that the composition of the cell membrane is adjusted to keep it above the phase transition temperature, and is regulated in response to environmental changes to maintain this condition \cite{Mouritsen_chapter}. In cases where this mechanism fails, cold shock damage takes place \cite{Drobnis}.

In this paper, we address the relationship between membrane microheterogeneity and anomalous subdiffusion in cell membranes by carrying out Monte Carlo simulations of two-component lipid membranes on experimentally relevant spatial scales ($\sim$1 $\mu$m) and time intervals ($\sim$1 s), with a special emphasis on interactions with the model membrane skeleton.

We demonstrate that (i) near-critical fluctuations in a free lipid membrane can lead to transient anomalous subdiffusion, (ii) phase separation in two-component (and, hence, multicomponent) lipid membranes can be strongly affected by interaction with the membrane skeleton, which, depending on the temperature and membrane composition, can either lead to precipitation of highly dynamic membrane domains (rafts), or prevent large-scale phase separation, and (iii) interaction with the membrane skeleton enhances anomalous subdiffusion and eventually leads to hop-diffusion of lipids. Thus, we construct a minimum realistic model for membrane rafts showing the features of both microscopic phase separation and anomalous subdiffusion.

The binary lipid system consisting of 1,2-di\-myr\-istoyl-\textit{sn}-glycero-3-phos\-pho\-cho\-line (DMPC) and 1,2-distearoyl-\textit{sn}-glycero-3-phosphocholine (DSPC) is chosen as a generic model of a
two-component membrane. The particular choice of this system is motivated by the fact that it is well studied both in vitro \cite{Shimshick, Mabrey, Lentz, Dijck, Wilkinson, Nibu, Sugar, Hac} and in silico (see \cite{Sugar, Hac, Heimburg_book} and refs. therein). Even though the exact temperature range of phase separation in this membrane is too high for most living organisms, the main qualitative conclusions of the study related to behavior of cell membranes are not affected by the particular thermodynamic parameters of the lipid mixture.

From the physical perspective, it is assumed that living organisms try to maintain the composition of cell membranes to keep them above the phase transition temperature \cite{Mouritsen_chapter}. Therefore, to be biologically relevant, in the present paper we mostly focus on the upper part of the phase diagram.

\section*{Materials and Methods}

\subsubsection*{Experimental}

The saturated phospholipids DMPC and DSPC with the melting temperatures 297 and 328 K were purchased from Avanti Polar Lipids (Alabaster, AL). Multilamellar vesicle (MLV) suspensions were obtained by rehydration of a dry lipid film. Excess heat capacity curves of DMPC/DSPC MLV suspensions in a 10 mM Hepes buffer, pH 7.4, were obtained using a VP-DSC calorimeter (MicroCal, Northampton, MA) at a scan rate of 2--3 K/h.

The empirical phase diagram based on the temperatures of the onset and completion of the phase transition determined from the scanning calorimetry data as described \cite{Sugar, Hac} are in agreement with previously reported results (see Supporting Material).

\subsubsection*{Monte Carlo simulations}

Our approach to lattice-based Monte Carlo (MC) simulations of a two-component membrane is generally similar to the one described previously \cite{Sugar, Hac}. To facilitate efficient simulations on experimentally relevant spatial scales ($\sim$1 $\mu$m) and time intervals ($\sim$1 s), where a particular type of lipid packing and fine molecular details should be of little importance, we further simplified the model and represented the membrane as a square lattice, each node of which represents a molecule of one of the two lipid types, which can be in either gel or fluid conformational state. An elementary MC step consists of two substeps: (i) an attempt to change the state of a randomly chosen lipid and (ii) an attempt of position exchange of a randomly chosen next-neighbor pair of lipids. As in \cite{Hac}, a rate function was introduced for the next-neighbor exchange step to ensure $\sim$40 times slower lipid diffusion in the gel phase. For an $L \times L$ lattice, an MC cycle consists of a chain of $L^2$ elementary MC steps. For every lipid composition and temperature studied, the membrane was first equilibrated, if required, and simulations were run for $(6-20) \times 10^{6}$ MC cycles to collect the necessary data. More details on the simulation procedure are given in the Supporting Material.

Simulations were carried out on an $L \times L = 600 \times 600$ ($400 \times 400$ when modeling the effects of the membrane skeleton) square lattice with periodic boundary conditions. By assuming the average lipid headgroup size of 0.8 nm, this corresponds to a membrane with an experimentally relevant size of $0.48 \times 0.48$ $\mu$m$^2$ ($0.32 \times 0.32$ $\mu$m$^2$). By comparing the DMPC diffusion coefficient obtained in our simulations of pure DMPC in the fluid phase at 304 K with its experimental values at the same temperature ($3 - 6$ $\mu$m$^2$/s \cite{Kuo, Vaz, Dolainsky}), we found that one MC cycle corresponds to $\sim$50 ns, and thus, our simulations cover processes on timescales up to $\sim$0.3--1 s.

\subsubsection*{Model for the membrane skeleton}

The membrane skeleton was modeled by a random Voronoi tessellation satisfying the periodic boundary conditions. For simulations with $L = 400$, random tessellations with $N = 36$ compartments were used, which gives the average linear size of the compartment $\ell = L N^{-1/2} = 66.6$ lattice units $\approx 53$ nm. The generated filament meshwork was projected onto the square lattice thereby creating a set of pixels representing the locations of the filaments. Each of these locations can be assigned to be a cytoskeleton pinning site. To mimic the membrane--cortical skeleton interaction, a simple rule inspired by experimental data on lipid interactions with transmembrane proteins \cite{Kinnunen} was followed: a lipid located at a filament pinning site is forced to assume the gel conformation with no explicit restrictions on its mobility. Thus, the pinning site does not present an obstacle for lipid diffusion: its effect on diffusion is indirect and takes place solely due to a lower lipid mobility in the gel-state local environment. The effect of the varying strength of the membrane-skeleton interaction was modeled by randomly assigning a fraction of filament position pixels to be filament pinning sites. Simulations were carried out with the filament pinning density set to 25\%, 50\%, and 100\%; in these cases the total number of pinning sites amounted to approximately 1\%,  2\%, and 4\% of the total membrane area.

Our choice of immobile pinning sites is justified by experimental observations demonstrating that band 3 and ankyrin strongly bound to the membrane skeleton show a very low diffusion coefficient of $\sim 10^{-4} - 10^{-3}$ $\mu$m$^2$/s over time intervals up to tens of seconds \cite{Tomishige, Cairo}.

It should be pointed out that the approach used in the present work to account for interactions of lipid molecules with membrane proteins is conceptually similar, though not identical, to the one used in several previous MC simulation-based studies \cite{Sperotto, Heimburg1996, Sabra, Gil}. In these works, it was assumed that proteins, preferentially wetted by one of the membrane phases, could freely diffuse in the membrane, which results in their accumulation in this phase and, in case of two-component membranes, in fact enhances large-scale phase separation. (The situation is more complicated in case of active proteins -- for details, see \cite{Gil} and refs. therein.) It was found out that, under certain conditions, interaction of a single-component membrane with small transmembrane proteins or peptides can even lead to emergence of a closed loop of gel--fluid coexistence with a lower critical point \cite{Zhang}. It should be pointed out, however, that these works focused only on structural properties of protein-loaded membranes, and did not address the issues of diffusion in the membrane.

What sets the present work apart from the above-mentioned studies, is that here we consider interaction of the membrane with the \textit{immobile} cytoskeleton and study its effects on phase separation and diffusion of lipids in the membrane.

\subsubsection*{Analysis of lipid diffusion data}

Positions of a small fraction of lipid molecules (150 for simulations with $L = 600$ and 50 for simulations with  $L = 400$, which amounted to 0.04\% and 0.03\%, respectively) were recorded, and the time- and ensemble-averaged mean-square displacement (MSD) was determined as follows:
\begin{equation}
\textit{MSD}(\tau) = \frac{1}{t_\text{max}-\tau} \sum_{t=1}^{t_\text{max}-\tau} \left\{ \left[x(t) - x(t+\tau) \right]^2
													                	              + \left[y(t) - y(t+\tau) \right]^2 \right\},
\label {Eq_MSD}
\end {equation}
where $\tau$, $t$, and $t_\text{max}$ are times measured in units of MC cycles;
here, $\tau$ denotes the time lag, and $t_{\text{max}}$ is the total length of the lipid molecule trajectory. No difference between time- and ensemble-averaged MSD (Eq. \ref{Eq_MSD}) and time-only- and ensemble-only-averaged MSDs beyond the statistical error level was observed.

In case of normal diffusion, the MSD grows in a linear fashion with time: $\textit{MSD}(\tau) = 4 D \tau$ in 2D, where $D$ is the translational diffusion coefficient. In case of anomalous subdiffusion, the MSD shows a slower sublinear power-law growth $\textit{MSD}(\tau) \sim \tau^{\beta}$ with $0<\beta<1$ (see, e.g., \cite{Metzler}). An alternative description of diffusion showing deviations from the normal behavior (also in cases where it cannot be described in terms of subdiffusion) can be provided using an effective time-dependent diffusion coefficient $D(\tau)$.

Therefore, to characterize the behavior of the MSD curves, the local exponent of the mean-square displacement
\begin{equation}
\beta_{\text{MSD}}(\tau) = d\log {\textit{MSD}(\tau)} /d \log \tau,
\label {Eq_beta_MSD}
\end {equation}
and the effective time-dependent diffusion coefficient
\begin{equation}
D(\tau) = \frac{1}{4} d \textit{MSD}(\tau)/d \tau
\label {Eq_D}
\end {equation}
were calculated.

\subsubsection*{Simulation of FCS experiments}

To simulate fluorescence correlation spectroscopy (FCS) \cite{FCS_review} measurements,
the tracked particles (as above, in the amount of 150 for simulations with $L = 600$ and 50 for simulations with  $L = 400$) were assumed to be fluorescent. Fluorescence intensity fluctuations $\delta F(t) = F(t) - \langle F \rangle$ about the mean intensity $\langle F \rangle$ in a 2D Gaussian detection spot $\exp(-2r^2/r_{0}^{2})$ were recorded, and their autocorrelation function $G(\tau) = \langle \delta F(t) \delta F( t+\tau ) \rangle / \langle F \rangle ^{2}$ was calculated. The detection spot size was set to $r_0 = 31$ lattice units $\approx 25$ nm, the size experimentally achievable using the stimulated emission depletion (STED) FCS technique \cite{Kastrup}. We additionally note that this detection spot is much smaller than the lattice size ($r_0/L \approx 0.05$ for $L = 600$ and $r_0/L \approx 0.08$ for $L = 400$), which allowed us to avoid artifacts in the fluorescence autocorrelation function and additionally ensured that our simulations are experimentally relevant.

FCS curves were averaged over nine different positions on the lattice; when studying the effects of the membrane skeleton, the results were additionally averaged over five random realizations of the filament network.

Simulated FCS curves were analyzed using the model
\begin{equation}
G(\tau) = G(0)/\left[ 1+( \tau /\tau _\text{D}) ^{\beta _{\text{FCS}}}\right].
\label {Eq_FCS}
\end {equation}

For $\beta_{\text{FCS}}=1$ this expression corresponds to normal diffusion, while for $0 < \beta_{\text{FCS}} < 1$ it provides a simple way to describe anomalous subdiffusion in FCS \cite{FCS_review}. In case of normal diffusion, $\tau_\text{D}$ is related to the diffusion coefficient of fluorescent particles $D$ and detection spot size $r_0$: $\tau_\text{D} = r_0^2/(4D)$.

Notice that since FCS is sensitive not only to the $\textit{MSD}(\tau)$, but also to the higher moments of the distribution of displacements of a diffusing particle, the connection between $G( \tau )$ and $\textit{MSD}(\tau)$ is straightforward only in case of a Gaussian distribution of displacements \cite{FCS_review}. As a result, in case of anomalous diffusion of particles, generally $\beta _{\text{FCS}} \neq  \beta_{\text{MSD}}$.

\section*{Results and Discussion}

\subsubsection*{Phase and component separation in the membrane}

\begin{figure}[!t]
\begin{center}
\includegraphics{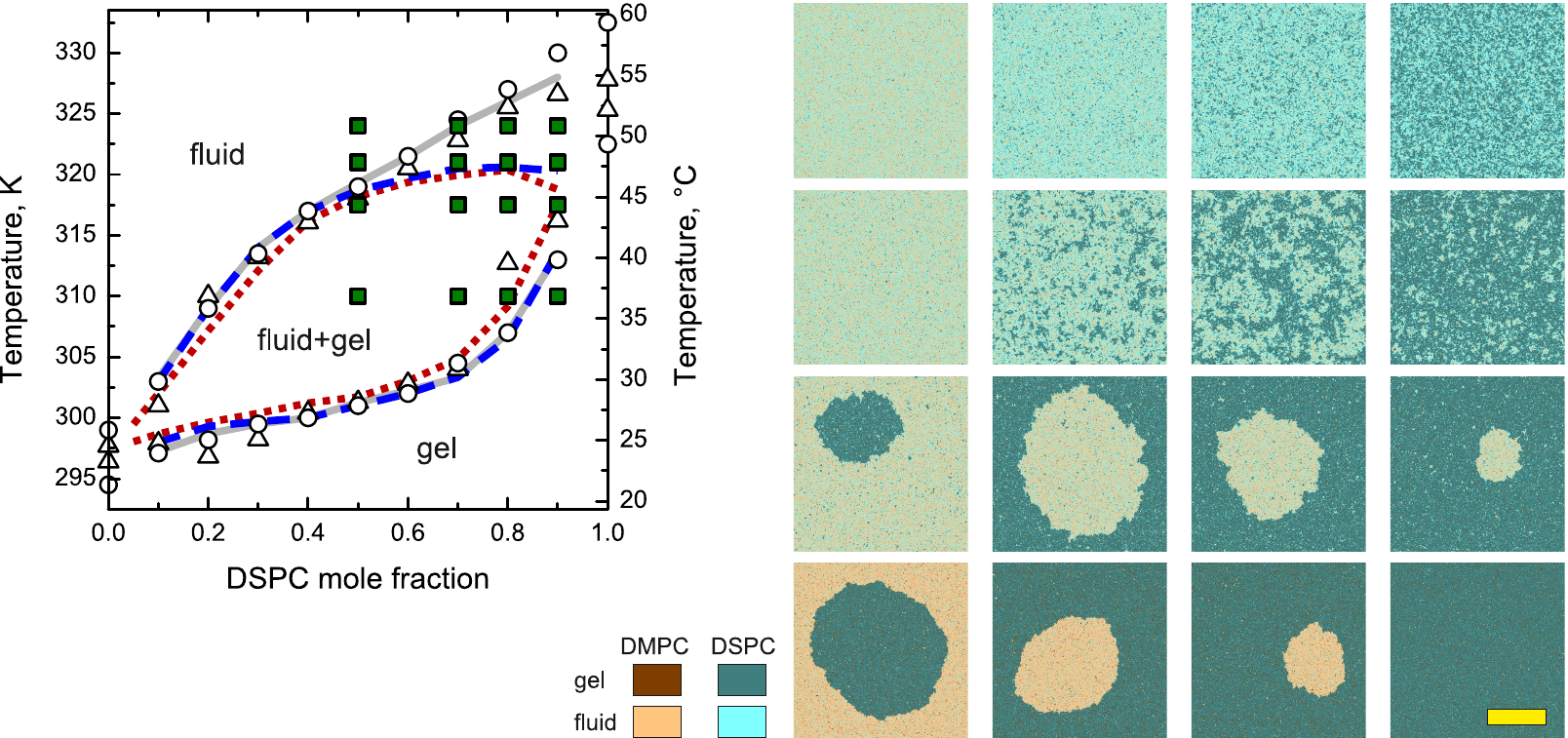}
\caption{Left-hand panel: Component and lipid state separation phase diagram of DMPC/DSPC lipid mixtures. Phase transition temperatures as determined from experimental ($\triangle$) and simulated ($\circ$) excess heat capacity curves. Lipid state spinodal (- - -), lipid state binodal (-- -- --), and lipid demixing curves (---). Right-hand panel: representative snapshots of equilibrium membrane configurations at temperatures and membrane compositions corresponding to filled squares in the left-hand panel. Lattice size: $600 \times 600$; scale bar: 200 lattice units $\approx 160$ nm.}
\label{Fig1}
\end{center}
\end{figure}

With appropriate tuning of lipid interaction parameters, the empirical heat capacity-based phase diagram obtained from our MC simulation is in agreement with our experimental data (Fig. \ref{Fig1}), as well as with previously published experimental and simulation data on the same lipid system (see Supporting Material). At the same time, the empirical phase diagram constructed on the basis of heat capacity data may differ from the real phase diagram of the system and thus not provide an insight into the microscopic structure and the dynamics of the membrane.

More details on the phase diagram can be obtained based on binodal and spinodal curves of the system. The binodal curves, also known as coexistence curves, are the boundaries of the region in the phase diagram in which the equilibrated system shows a complete separation of the two phases. The spinodal encloses the region where the mixture is unstable with respect to small local fluctuations of the composition and always lies inside the area enclosed by the binodal, with the exception of a critical point, where the binodal and spinodal touch.

To reconstruct the binodals and spinodals of the DMPC/DSPC lipid mixture, we analyzed the static structure factors \cite{Fisher} for lipids $S_\text{L}(k)$ and lipid states $S_\text{S}(k)$ (see Supporting Material), reflecting the character of spatial fluctuations of the membrane composition and state.

We found that outside the phase coexistence region, the Ornstein--Zernike (OZ) approximation \cite{Fisher} $S^{\text{OZ}}(k) = S(0)/[ 1 + ( k \xi )^{2}] $, where $k$ is a wavenumber, and $\xi$ is a correlation length, provided an excellent description of the lipid-state structure factors $S_{\text{S}}(k)$. The temperature dependences of $S_{\text{S}}(0)$ and $\xi_{\text{S}}$ were used to estimate spinodal and binodal curves. In particular, the spinodals were determined by extrapolating $1/S_{\text{S}}(0)$ dependence to zero crossing \cite{Fisher, Strobl}, and binodals were determined from the condition $d(1/\xi_{\text{S}})/dT = 0$ \cite{Strobl}. Interestingly, in some regions of the phase diagram, the binodal deviates quite strongly from the phase coexistence boundary estimated from the excess heat capacity data (Fig. \ref{Fig1}). This discrepancy indeed shows that the analysis of heat capacity data does not necessarily recover the real phase diagram of the system. Here, this behavior reflects the continuous character of the phase transition in the membrane, whose proper description requires a combined component and state phase diagram \cite{Michonova-Alexova}.

What process is reflected by the excess heat capacity curves where they fail to describe the fluid--gel phase separation? An analysis of the structure factors $S_{\text{L}}(k)$ for the lipid species helps to answer this question. It appears that, generally, two OZ components are required to describe these data: $S_{\text{L}}(k) = S_{\text{L}1}^{\text{OZ}}(k) + S_{\text{L}2}^{\text{OZ}}(k)$. Parameters of component 1 only weakly depend on the temperature and reflect demixing of lipids in the same state. Parameters of the second component $S_{\text{L}2}(0)$ and $\xi_{\text{L}2}$ show strong temperature dependences similar to those of $S_{\text{S}}(0)$ and $\xi_{\text{S}}$, and thus describe appearance of dynamic microscopic domains, which cannot be treated as distinct thermodynamically stable phases \cite{Heimburg_book}. We therefore define the lipid demixing curves as a set of points on the phase diagram satisfying the condition $S_{\text{L}1}(0) = S_{\text{L}2}(0)$. The fact that the lipid demixing curves are in excellent agreement with the calorimetry-based empirical phase diagrams, supports the above reasoning, as do the snapshots presented in Fig. \ref{Fig1}. Notice that in this region the membrane undergoes near-critical fluctuations. On the other hand, in the coexistence region, equilibrated membranes show complete phase separation and form large-scale lipid domains (Fig. \ref{Fig1}).

The above analysis of the structure factors $S_{\text{L}}(k)$ and $S_{\text{S}}(k)$ and equilibrated membrane snapshots shows that in the part of the phase diagram where the lipid demixing curve closely approximates the binodal, the phase transition from the fluid state to the fluid--gel coexistence has a quasi-abrupt character. On the other hand, in the region where the lipid demixing curve strongly deviates from the binodal, the membrane shows characteristic near-critical fluctuations (Fig. \ref{Fig1}), and the transition becomes continuous as the system approaches the critical point (DMPC/DSPC $\approx$ 20:80, $T_{\text{c}} = 320.5 \pm 0.2$ K -- see \cite{NJP} for details). Remarkably, this behavior is qualitatively similar to recent experimental observations on three-component lipid membranes \cite{Veatch}: depending on the membrane composition, the transition to the two-phase coexistence takes place either in an abrupt manner -- when the membrane does not pass through a critical point, or via critical fluctuations -- when the membrane does pass through a critical point. A more detailed consideration of the phase diagram of the DMPC/DSPC system based on MC simulations will be published elsewhere \cite{NJP}.

In the fluid--gel phase coexistence region, large-scale phase separation takes place. Since no explicit or implicit penalties are imposed \cite{Baumgart2003, Semrau, Ursell} on the domain size in our model, and domain growth is driven by minimization of the line tension energy, phase separation results in formation of a single circular-shaped domain of the minority phase (Fig. \ref{Fig1}).

\subsubsection*{Effect of the cytoskeleton on phase separation in the membrane}

\begin{figure}[!t]
\begin{center}
\includegraphics{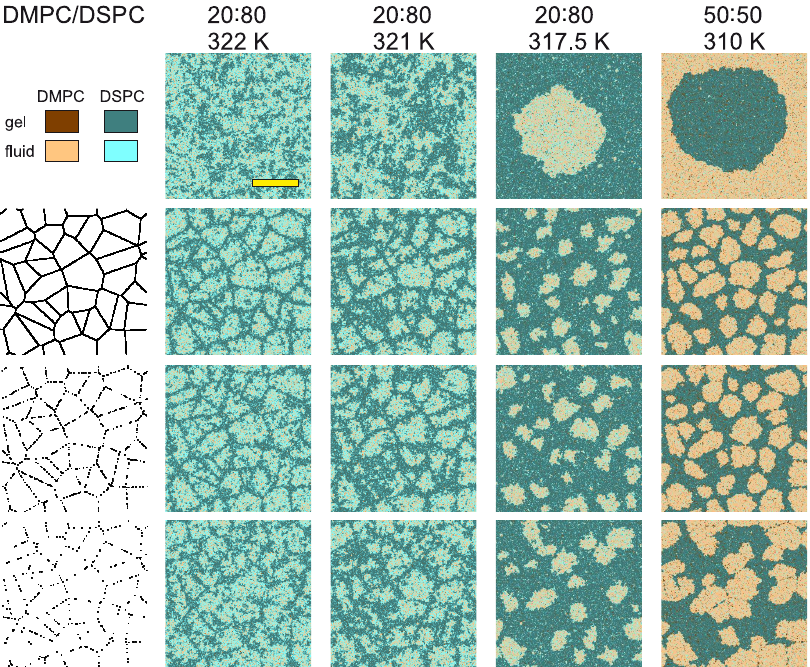}
\caption{Effect of the membrane skeleton on the phase separation in a DMPC/DSPC membrane. Representative snapshots of membrane configurations are shown for the free membrane (\textit{first row}) and membrane interacting with a network of filaments at 100$\%$ (\textit{second row}), 50$\%$ (\textit{third row}) and 25$\%$ pinning density (\textit{fourth row}). Snapshots for the free membrane, as well as for the membrane interacting with filaments at $T = 321$ and 322 K, represent fully equilibrated configurations; snapshots for the membrane interacting with filaments at $T = 310$ and 317.5 K correspond to equilibration time of $6 \times 10^6$ MC cycles (see text for discussion). For presentation purposes, the filaments are drawn thicker than they are in reality. Lattice size: $400 \times 400$; scale bar: 125 lattice units $\approx 100$ nm.}
\label{Fig2}
\end{center}
\end{figure}

In the region of near-critical fluctuations the membrane is expected to be very sensitive to external perturbations, including the interaction with the membrane skeleton. It appears that in the vicinity of the critical point the interaction with the membrane skeleton leads to immediate condensation of the gel phase on the skeleton filaments, and formation of membrane domains, in a good agreement with results of experiments on lipid bilayers with a reconstituted actin skeleton \cite{Liu}. These domains dynamically change their shape, but nevertheless stay pinned to the filaments (Fig. \ref{Fig2}). Notice that the effect of the filaments is remarkably robust with respect to the filament pinning density. We strongly believe that these domains represent the minimal model of membrane rafts \cite{Pike}. The minimum character of this model stems from the fact that in the present scenario domain formation is thermodynamically driven, and does not require any active processes like chemical cross-linking of membrane components or lipid recycling.

\begin{figure}[!t]
\begin{center}
\includegraphics{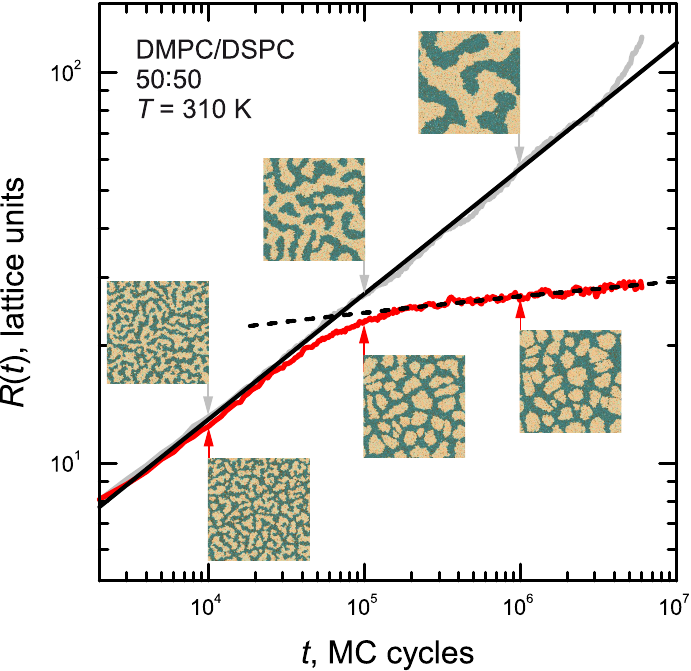}
\caption{Effect of the membrane skeleton on the the domain growth in DMPC/DSPC 50:50 membrane abruptly cooled from the all-fluid state down to $T = 310$ K in the fluid--gel phase coexistence region. Kinetics of domain growth when the membrane is free (\textit{upper curve}) and in the presence of membrane skeleton with the filaments pinning density of 50$\%$ (\textit{lower curve}). Black solid line shows the power law dependence $R(t) \sim t^n$ with $n = 0.32$; dashed line shows the stage of the slow logarithmic growth $R(t) \sim \ln t$ in the presence of membrane skeleton. Representative membrane configurations obtained in our MC simulations at time instants $10^4$, $10^5$, and $10^6$ MC cycles are shown at the corresponding curves. Lattice size: $400 \times 400$.}
\label{Fig3}
\end{center}
\end{figure}

If the membrane is abruptly cooled down from the all-fluid state to a temperature in the fluid--gel coexistence region of the phase diagram, phase separation takes place, and domains of the fluid and gel phase start to nucleate and coarsen with time. In a free membrane, domains grow according to the power law $R(t) \sim t^n$ with the growth exponent $n$  depending on the particular domain growth mechanism \cite{Furukawa}. In our simulations for a free membrane, depending on the lipid composition and temperature, $n$ takes values from $1/4$ to $1/3$ (Fig. \ref{Fig3}), consistent with general expectations for the domain growth in 2D \cite{Furukawa} and in agreement with experimental results (see, e.g., \cite{Jensen}) (for a more detailed discussion, see \cite{NJP}).

In a finite-size system with a linear dimension $L$, domain coarsening eventually stops, and a single circular-shaped domain is produced (Fig. \ref{Fig1}) with the radius $R_\infty \simeq (X/\pi)^{1/2}L$, where $0 < X < 1/2$ is the fraction of the minority phase.

Remarkably, we found that the presence of the membrane skeleton strongly inhibits or even eventually prevents large-scale phase separation in the phase coexistence region (Fig. \ref{Fig2}). As a result, the radius of the membrane domains in this case is largely determined by the characteristic compartment radius $R_{\text{comp}} \simeq \ell/2$ of the filament network. The way we account for interaction of lipid molecules with filaments at their pinning sites is similar to a theoretical model (the random-field Ising model \cite{RFIM}) implying the presence of static random position-dependent perturbations in a 2D system of spins. This model predicts that the initial power-law domain growth $R(t) \sim t^n$ is strongly slowed down at intermediate stages, and crosses over to an extremely slow logarithmic growth $R(t) \sim \log t$; eventually the domains are expected to reach the perturbation strength-dependent maximum size $R' < R_\infty$ in an exponential time to create an equilibrium disordered state \cite{RFIM_theory, RFIM_simulations}. In our system, the perturbation strength is determined by the filament pinning density and the average compartment radius $R_{\text{comp}}$. Therefore, if the interaction of the membrane with the cytoskeleton is strong enough, and ${R_\infty}^2 \gg R_{\text{comp}}^2$, one can expect that the domain growth stops when domains reach the characteristic size $R_{\text{comp}} \lesssim R' < R_\infty$.

We indeed observed an extreme slowing down of the domain growth in the presence of the membrane skeleton and cross-over to the slow logarithmic growth (Fig. \ref{Fig3}). As is evident from the Figure, at the end of a simulation run of $6 \times 10^6$ MC cycles, the domain sizes are indeed $R(t) \sim R_{\text{comp}}$. Although some growth is still observed at this stage, it is so slow, that at present it is unclear, whether the system will evolve toward a disordered equilibrium state featuring a number of small domains, as suggested in \cite{RFIM_theory, RFIM_simulations}, or the slow logarithmic growth will continue further to produce eventually (after an extremely long time) a single domain with the size $R_\infty$. What is clear, though, that the time required for complete equilibration is so long, that from the practical viewpoint one can state that indeed the presence of the membrane skeleton prevents large-scale phase separation in the membrane. This becomes especially clear if one takes into account that a membrane in a live cell is not in the equilibrium state, and a number of other processes affecting the membrane state and composition, e.g., lipid recycling, take place in parallel on different spatial and time scales.

Thus, the cytoskeleton-induced inhibition of large-scale phase separation can serve as one of the possible explanations why micrometer-scale membrane domains are observed in giant plasma membrane vesicles \cite{Plasma_membrane_vesicles}, but not in cell membranes.

In addition, our observations suggest that the established cryoprotective role played by the cytoskeleton \cite{Clark} may consist in delaying phase separation-induced cold shock damage of living cells.

\subsubsection*{Diffusion of lipids in the free membrane}

\begin{figure}[!t]
\begin{center}
\includegraphics{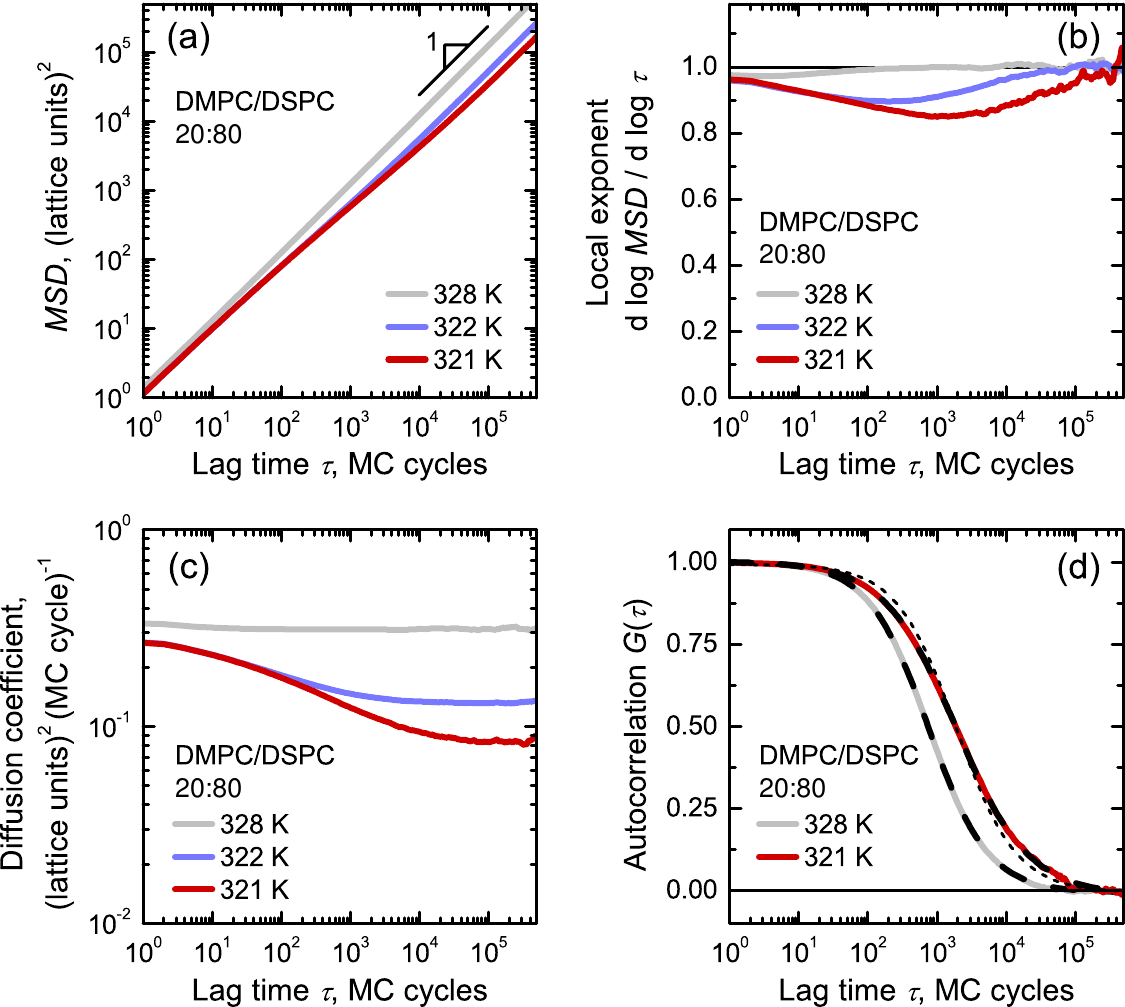}
\caption{Effect of the proximity to the phase transition on diffusion of DMPC lipids in a DMPC/DSPC 20:80 membrane. Mean-square displacement $\textit{MSD}(\tau)$ (a), local exponent $\beta_{\text{MSD}}(\tau)$ (b), time-dependent diffusion coefficient $D(\tau)$ (c), and FCS autocorrelation $G(\tau)/G(0)$ (d). Panels (a) -- (c) top to bottom: $T = 328$, 322, and 321 K. Panel (d): $T = 328$ (\textit{left}) and 321 K (\textit{right}). For clarity, data for 322 K are omitted in panel (d). Panel (d) additionally shows fits to the FCS diffusion model Eq. (\ref{Eq_FCS}) giving $\beta_{\text{FCS}} = 1.01$ at 328 K and $\beta_{\text{FCS}} = 0.86$ at 321 K (-- -- --). For comparison, a fit of 321 K data with fixed $\beta_{\text{FCS}} = 1.0$ is shown (- - -).}
\label{Fig4}
\end{center}
\end{figure}

At higher temperatures, away from the phase transition and the near-critical fluctuations region, the membrane is in the homogeneous all-fluid state, and, not unexpectedly, lipid diffusion is normal (Fig. \ref{Fig4}).

This picture changes upon approaching the critical point in the region of near-critical fluctuations where  interpenetrating fluctuating domains form in the membrane. The analysis of the immediate environment of the DMPC and DSPC lipids shows that, while the DSPC lipid does not have a pronounced preference for the state of its local environment and therefore is evenly distributed between fluid and gel domains, the DMPC lipid shows a strong preference for the fluid local environment and is thus predominantly partitioned into fluid domains. Under these conditions, the DSPC lipid shows no significant deviation from the normal diffusion behavior (data not shown). By contrast, quite unexpectedly, the DMPC lipid was found to demonstrate very pronounced anomalous subdiffusion (Fig. \ref{Fig4}).

At a first glance, this subdiffusive behavior is rather surprising, especially in the light of the seminal work by Kawasaki \cite{Kawasaki} demonstrating that the diffusion coefficient of a two-component mixture does not vanish upon approaching the critical point, which implies normal diffusion, at least in the long-time asymptotic regime. A closer look at the data shows, however, that the subdiffusive behavior is in fact transient, though it covers several orders of magnitude in time, and at longer times the cross-over to the normal diffusion takes place. One can also notice that at very short time lags, diffusion is normal as well. This behavior is also evident from the plots of the local exponent $\beta_{\text{MSD}}(\tau)$ and effective time-dependent diffusion coefficient $D(\tau)$ extracted from the $\textit{MSD}(\tau)$ dependences (Fig. \ref{Fig4}b, c).

How this cross-over from normal diffusion to a subdiffusive behavior and back to normal diffusion at long times can be explained? By recalling that the DMPC lipid is preferentially partitioned into fluid domains and having a look at the corresponding membrane snapshots (Fig. \ref{Fig1}), we realize that DMPC molecules diffuse on a dynamically rearranging fractal-like fluid phase pattern. Therefore, at early times, when the DMPC molecule explores its immediate environment, normal diffusion should take place. Later, though at times shorter than a characteristic time of rearrangement of this dynamic fractal structure, the motion of DMPC lipid molecules is subdiffusive, similar to what is expected in case of diffusion on fractal-like percolation clusters \cite{ben-Avraham_Havlin}. At much longer times, evolution of these clusters leads to dynamic percolation behavior (see, e.g., \cite{dynamic_percolation}), which results in a crossover from the subdiffusive motion to normal diffusion. In contrast to diffusion on static percolation clusters, where cross-over to normal diffusion occurs only above the percolation transition \cite{ben-Avraham_Havlin}, in the case of lipid diffusion in a near-critical membrane, cross-over to normal diffusion will always take place irrespectively of whether the fluid phase is above or below the percolation threshold.

This subdiffusive behavior is also clearly observed in our simulations of FCS experiments (Fig. \ref{Fig4}d). Upon a closer approach to the critical point of the system, the autocorrelation functions of fluorescence intensity fluctuations $G(\tau)$ progressively stronger deviate from the dependence expected for normal diffusion (i.e., Eq. (\ref{Eq_FCS}) with $\beta_{\text{FCS}} = 1$). A good description of $G(\tau)$ data in this case can only be obtained with $\beta_{\text{FCS}} < 1$. For example, we find that for DMPC/DSPC 20:80 mixture at $T = 321.0$ K the best fit is obtained with $\beta_{\text{FCS}} = 0.86$ (Fig. \ref{Fig4}d), and a smaller $\beta_{\text{FCS}} = 0.79$ is required to fit $G(\tau)$ at $T = 320.7$ K, which is closer to the critical point (data not shown).

A new important result of the present work is that transient anomalous subdiffusion spanning several orders of magnitude in time can be observed in a system close to its critical point. We emphasize that the appearance of the transient subdiffusive behavior in the region of near-critical fluctuations is not related to any specific properties of the system and should be observed close to criticality in various systems independent of their origin, including, of course, multicomponent lipid membranes.

It is well known that phase separation in lipid bilayers produces domains with typical sizes ranging from a few to several tens of micrometers (provided that there are no restrictions on the domain growth). In agreement with that, we observe in our simulations that within the coexistence region an equilibrated membrane always shows complete phase separation resulting in a single circular-shaped domain of the minority phase. Therefore, in this case, the only reasonable way to carry out FCS measurements requires parking the detection spot into the bulk of the majority phase away from the inter-phase boundary, or into the center of the minority phase domain. Exactly this approach is used in experimental FCS studies on membranes showing large-scale phase separation (see, e.g., \cite{Bacia, Kahya, Chiantia}). Not unexpectedly, this approach results in normal diffusion in both phases with phase-dependent diffusion coefficients. This behavior is also observed in our simulations (data not shown).

In contrast to that and quite surprisingly, an MC simulation study of Hac et al. \cite{Hac} reported FCS curves strongly deviating from the normal diffusion model exactly in the phase coexistence region, i.e., where large-scale phase separation takes place. These results were attributed by the authors of \cite{Hac} as resulting from subdiffusive motion of lipids, which is in contradiction with the macroscopic phase separation in the system. This seeming contradiction is resolved upon careful examination of the approach to simulations in \cite{Hac}. There, the diameter of the FCS detection spot was approximately equal to the simulation box size. In this case, the FCS detection spot covers effectively the whole simulated membrane, and, even in the presence of large-scale phase separation, motion of lipids in both fluid and gel phases contributes to the FCS results, which explains deviations of FCS curves from the normal diffusion model observed in \cite{Hac} in the fluid--gel phase coexistence region and rule out the interpretation of these results in terms of subdiffusion. Moreover, the situation simulated in \cite{Hac} is very unlikely for experiments on single lipid bilayers within the phase coexistence region, where membrane domains typically have radii of a few to several tens of micrometers, whereas the typical FCS detection spot size is $\sim 200$ nm for the standard FCS \cite{FCS_review}, and down to 25 nm for STED FCS \cite{Kastrup}.

Here we should also mention a previous lattice-based simulation study \cite{Sugar2005main} addressing the character of lateral diffusion of lipid molecules in DMPC/DSPC 50:50 membranes. There, in agreement with our results, diffusion is normal at $T > 320$ K, i.e. in the all-fluid membrane state (cf. Fig. \ref{Fig1}). Deviations from the normal diffusion for the DMPC/DSPC 50:50 mixture found in \cite{Sugar2005main} within the temperature range $T = 300 - 320$ K (i.e. exactly where the large-scale phase separation for this lipid composition takes place -- see Fig. \ref{Fig1}) were interpreted in terms of subdiffusive motion of lipids. As in \cite{Hac}, the spatial scales on which deviations from normal diffusion were observed in \cite{Sugar2005main} approach the system size. Hopping of lipid molecules between the macroscopic gel and fluid domains definitely leads to deviations from normal diffusion at these large observation scales, although these deviations can hardly be interpreted in terms of anomalous subdiffusion.

We emphasize that our results presented in this paper unambiguously show that local component and phase fluctuations in a near-critical membrane lead to the transient anomalous subdiffusion spanning several orders of magnitude in time. Under suitable experimental conditions, we believe, this subdiffusive behavior can be observed experimentally using, e.g., the (STED) FCS technique.

\subsubsection*{Effects of membrane--cytoskeleton interaction on lipid diffusion}

\begin{figure}[!t]
\begin{center}
\includegraphics{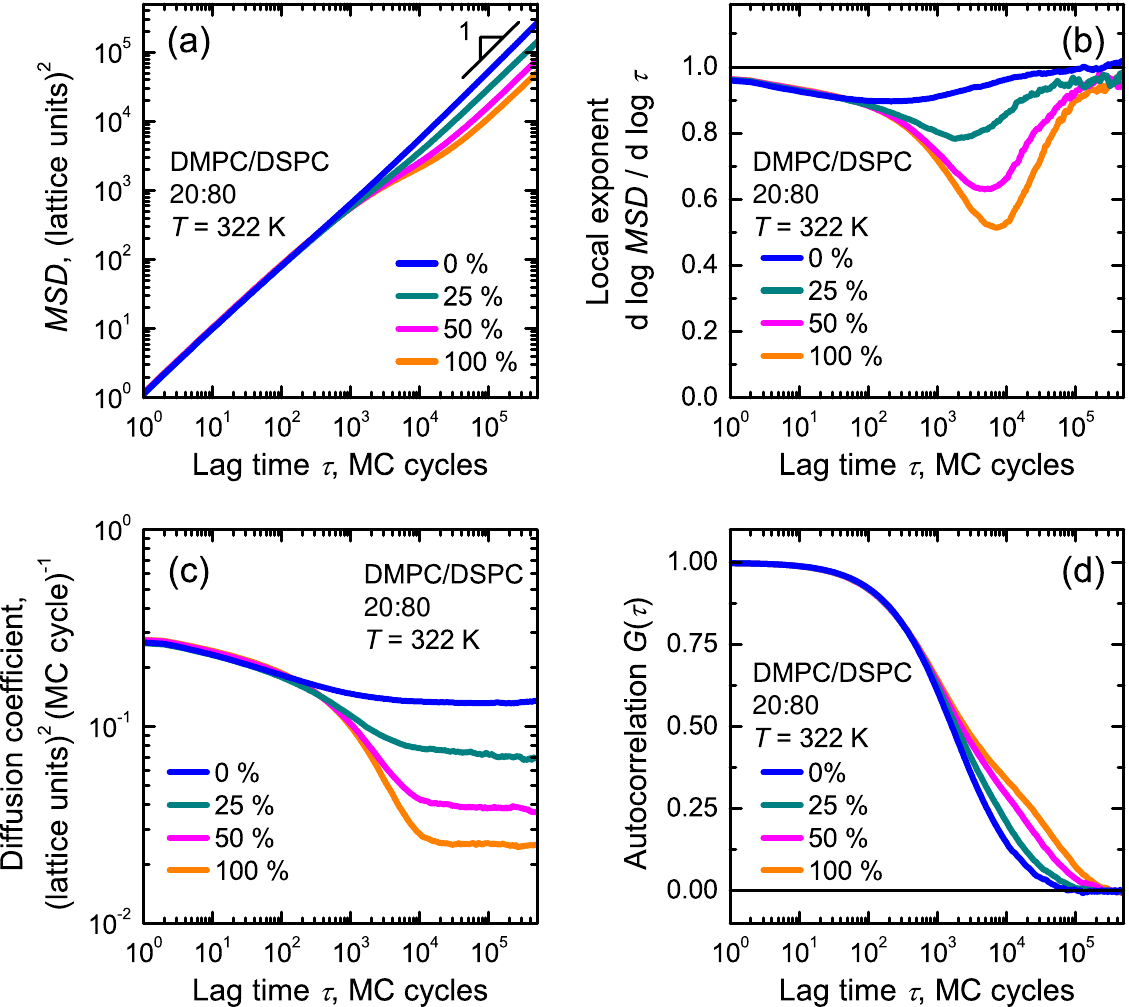}
\caption{Effect of the membrane skeleton on diffusion of DMPC lipids in a DMPC/DSPC 20:80 membrane at 322 K. Mean-square displacement $\textit{MSD}(\tau)$ (a), local exponent $\beta_{\text{MSD}}(\tau)$ (b), time-dependent diffusion coefficient $D(\tau)$ (c), and FCS autocorrelation $G(\tau)/G(0)$ (d). Panels (a) -- (c) top to bottom, and panel (d) left to right: filament pinning density equals 0$\%$ (free membrane), 25$\%$, 50$\%$, and 100$\%$.}
\label{Fig5}
\end{center}
\end{figure}

Interaction with the membrane skeleton strongly affects the character of diffusion in the region of near-critical fluctuations (Fig. \ref{Fig5}). With increasing the filament pinning density, the interaction of the membrane with filaments becomes more pronounced: it first enhances the anomalous diffusion and eventually leads to hop-diffusion of lipid molecules. In the presence of filaments, the faster diffusion process clearly corresponds to Brownian motion within compartments defined by the membrane skeleton, whereas the slower diffusion is due to hopping between the compartments, and therefore it strongly depends on the filament pinning density. Notice that gel phase condensation at the membrane skeleton substantially increases the effective thickness of filaments and thus enhances their influence on lipid diffusion.

Remarkably, in this case our results for the mean-square displacement $\textit{MSD}(\tau)$ (Fig. \ref{Fig5}a) and effective time-dependent diffusion coefficient $D(\tau)$ (Fig. \ref{Fig5}c) are in good qualitative agreement with results of single-molecule tracking experiments in cell membranes \cite{Murase}, where results were interpreted in terms of a picket-fence model. Our $\textit{MSD}(\tau)$ data also qualitatively agree with a theoretical model for diffusion on an infinite periodic square meshwork of penetrable barriers \cite{Wieser}, except for in our case the short-time motion of a molecule within a compartment is subdiffusive due to near-critical fluctuations. We again emphasize that even a relatively low pinning density of the cytoskeleton filaments leads to appearance of rather strong barriers for diffusing lipids due to filament-induced local condensation of the gel phase.

Our simulations of STED FCS experiments with the experimentally achievable detection spot size \cite{Kastrup, Eggeling} show (Fig. \ref{Fig5}d) that, while at early times all FCS curves exhibit behavior characteristic of subdiffusion, interaction with the membrane skeleton leads to appearance of a second slow diffusion component. The contribution of this slow component increases with the filament pinning density. Clearly, this component is related to the large-scale diffusion that involves crossing the filament-induced barriers on the membrane. This two-component character of $G(\tau)$ is in qualitative agreement with the recent STED FCS measurements of lipid diffusion in cell membranes \cite{Eggeling}. It would be very interesting to follow the dependence of the FCS autocorrelation functions and the respective apparent diffusion coefficients on the spatial scale using the variable detection spot FCS technique \cite{Masuda, Wawreziniek}, as has been done in \cite{Eggeling}. This work is currently in progress.

Here, we point out once again that in our model the pinning sites do not represent obstacles for diffusing lipids, and affect lipid diffusion only indirectly, via the lower lipid mobility in the local gel-state environment (see Materials and Methods). We anticipate that in case where pinning sites would be presented by immobile particles preferentially wetted by gel-phase lipids, the effects of the membrane--filament interaction on diffusion of lipids should become even more pronounced. This issue will be addressed in our future work.

\section*{Conclusion}
In conclusion, in this paper, we carried out Monte Carlo simulations of model two-component lipid membranes on experimentally relevant spatial scales and time intervals. This allowed us to better understand the details of the dynamics of phase and component separation, and by this means to address the relationship between membrane microheterogeneity and anomalous subdiffusion in cell membranes.

We observed that interaction of the near-critical membrane with the cytoskeleton strongly affects phase separation: for the membrane in the state of near-critical fluctuations it leads to local phase separation and formation of dynamic domains with a size of a few tens of nanometers, which conform to the current understanding of membrane rafts \cite{Pike}.

We find that in a membrane showing near-critical fluctuations lipids show subdiffusive behavior covering several orders of magnitude in time. The interaction of the membrane with the cytoskeleton enhances subdiffusion and eventually leads to hop-diffusion of lipids.

In the fluid--gel phase coexistence region of the phase diagram, interaction of the membrane with the cytoskeleton is found to tremendously slow down large scale phase separation in the membrane, which may explain the established connection between the cytoskeleton and cold shock resistance of organisms \cite{Clark}.

The concepts of lipid rafts \cite{Simons} and cytoskeleton-related picket fence \cite{Fujiwara} are frequently discussed as the alternative viewpoints on the origin of anomalous diffusion in cell membranes. In our paper, we bring these two concepts together to show that not only they do not contradict one another, but rather work in synergy, resulting in formation of cytoskeleton-induced dynamic lipid domains in the near-critical membrane. By this means we construct what, we believe, is a minimum raft model, since the domain formation is driven solely by thermodynamics and does not require either chemical cross-linking of membrane components, or lipid recycling.

\section*{Acknowledgements}
We acknowledge inspiring discussions with H. Rigneault and C. Favard at the early stage of the project.

The work was supported by the Deutsche Forschungsgemeinschaft via Research Group FOR 877.

\newpage
\thispagestyle{empty}
\vfill

\makeatletter \renewcommand\@biblabel[1]{[S#1]} \makeatother
\renewcommand\citenumfont[1]{S#1}

\makeatletter \renewcommand{\thepage}{S\@arabic\c@page}
\makeatletter \renewcommand{\thefigure}{S\@arabic\c@figure}
\makeatletter \renewcommand{\thetable}{S\@arabic\c@table}
\makeatletter \renewcommand{\theequation}{S\@arabic\c@equation}

\setcounter{page}{1}
\setcounter{figure}{0}
\setcounter{table}{0}
\setcounter{equation}{0}

\begin{center}\textbf{\Large Near-Critical Fluctuations and Cytoskeleton-Assisted Phase Separation Lead to Subdiffusion in Cell Membranes}\end{center}

\vfill

\begin{center}
Jens Ehrig, Eugene P. Petrov,\footnote[1]{Corresponding author: petrov@biotec.tu-dresden.de} and Petra Schwille\par
\end{center}

\begin{center}
\textit{Biophysics, BIOTEC, Technische Universit\"{a}t Dresden, Tatzberg
47/49, 01307 Dresden, Germany\\}
\end{center}

\vfill
\begin{center}
{\Large\textbf{-- Supporting Material --}}
\end{center}

\vspace{70mm}

\newpage

\subsection*{Empirical phase diagram based on differential scanning calorimetry data}
A single-component membrane undergoes a phase transition from the gel state to the fluid state at its melting temperature. On the other hand, for a two-component lipid membrane the transition from the all-gel state to the all-fluid state is not immediate. (In fact, the two-component membrane undergoes two phase transitions: one from the gel state to the fluid--gel coexistence and, at a higher temperature, another one from the fluid--gel coexistence to the fluid state.) Therefore, from the practical point of view, one can speak about a broadened gel--fluid transition in a two-lipid system (compared to a single-lipid system) \cite{Shimshick1973, Wilkinson1979, Sugar1999}. This broadened transition can be characterized by the onset and completion temperatures, which can be determined experimentally. By plotting the experimentally determined onset and completion temperatures as a function of the membrane composition, one obtains an empirical experimental phase diagram of the system.

To determine the onset and completion temperatures from the experimental DSC data, the empirical tangent method was used. The outer slopes of the $C(T)$ profiles of a range of compositions (DMPC/DSPC = 0/100, 10/90, 20/80, ..., 90/10, 100/0) were fit with straight lines passing through the corresponding inflection points. The onset and completion temperatures in this case are defined as the respective intersections of the tangential lines with the zero-line.

\begin{figure}[!t]
\begin{center}
\includegraphics{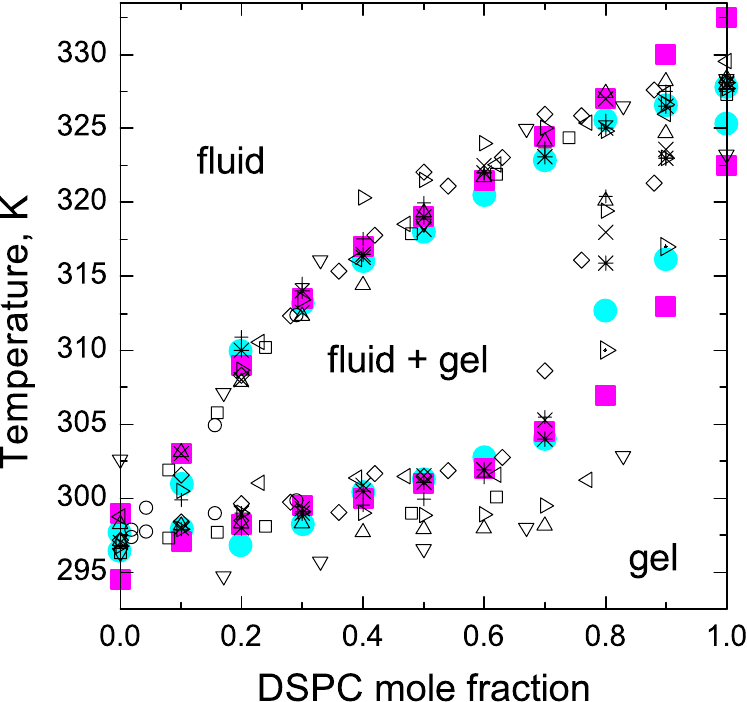}
\caption{Empirical phase diagram of DMPC/DSPC constructed from excess heat capacity curves measured experimentally on multilamellar vesicle suspensions and calculated from MC simulation data. Experimental data: present work ({\Large $\bullet$}), Ref. \cite{Shimshick1973}~({\footnotesize $\square$}), Ref. \cite{Wilkinson1979} ($\circ$), Ref. \cite{Mabrey1976}~($\diamond$), Ref. \cite{Lentz1976} ($\triangledown$), Ref. \cite{Dijck1977} ($\vartriangle$), Ref. \cite{Nibu1995}~($\triangleleft$), Refs. \cite{Sugar1999, Sugar2000}~($\triangleright$), our reanalysis of experimental $C(T)$ data for DMPC/DSPC 20/80 and 10/90 from \cite{Sugar2000} ($\triangleright$\hspace{-1.9mm}$\cdot$\hspace{0.5mm}), Ref. \cite{Hac2005}~($\times$). Monte Carlo simulation data: present work ({\footnotesize $\blacksquare$}), Refs. \cite{Sugar1999, Sugar2000} ($+$), Ref. \cite{Hac2005} ($\ast$).}
\label{FigS1}
\end{center}
\end{figure}

The experimental phase diagram of the DMPC/DSPC lipid mixture obtained in the present study is in a good agreement with the ones published earlier by other groups \cite{Shimshick1973, Wilkinson1979, Sugar1999, Mabrey1976, Lentz1976, Dijck1977, Nibu1995, Sugar2000, Hac2005} (Fig. \ref{FigS1}).

We remark here that evaluation of the onset and completion temperatures is not always unambiguous. In particular, our reanalysis of the experimental $C(T)$ data from \cite{Sugar2000} for DMPC/DSPC 20/80 and 10/90 mixtures gives onset temperatures lower than those reported in \cite{Sugar2000} and very close to the ones we obtained from our data (Fig. \ref{FigS1}).

\subsection*{Monte Carlo simulations}

The fundamental step of the MC simulation consists of two sub-steps. In the first sub-step, an attempt is made to change the state of a randomly chosen lipid (from gel to fluid or vice versa). The second sub-step is an attempt to exchange the positions of a randomly chosen pair of next neighbors on the lattice.

For each sub-step the change in the Gibbs free energy
\begin{equation}
\begin{split}
\Delta G =  & \Delta N^{\rm F}_1 \left(\Delta E_1 -T\Delta S_1 \right)
							+ \Delta N^{\rm F}_2 \left(\Delta E_2 -T\Delta S_2 \right)
							+ \Delta N^{\rm GF}_{11}w^{\rm GF}_{11}
		  					+ \Delta N^{\rm GF}_{22}w^{\rm GF}_{22} \\
			& + \Delta N^{\rm GG}_{12}w^{\rm GG}_{12}
		 					+ \Delta N^{\rm FF}_{12}w^{\rm FF}_{12}
		 					+ \Delta N^{\rm GF}_{12}w^{\rm GF}_{12}
		 					+ \Delta N^{\rm GF}_{21}w^{\rm GF}_{21},
\label{eq:Gibbs}
\end{split}
\end{equation}
is calculated. Here, $\Delta E_i$ and $\Delta S_i$ are the changes in the internal energy and entropy of a molecule of lipid $i$ when it switches its state from gel to fluid, and $w^{mn}_{ij}$ are the next-neighbor interaction parameters of lipids $i$ and $j$ being in states $n$ and $m$, respectively. In the simulation, the attempts of the state change and next-neighbor exchange are accepted with probability $p = 1$ for $\Delta G < 0$ and $p = \exp \left(-\Delta G/RT\right)$ for $\Delta G \geq 0$.

For an $L \times L$ lattice, one MC cycle consists of a chain of $L^2$ elementary MC steps. After each cycle, the enthalpy of the lattice is calculated as follows:
\begin{equation}
\begin{split}
H = 	& N^{\rm F}_1 \Delta E_1
					+ N^{\rm F}_2 \Delta E_2
					+ N^{\rm GF}_{11}w^{\rm GF}_{11}
					+ N^{\rm GF}_{22}w^{\rm GF}_{22}\\
		& + N^{\rm GG}_{12}w^{\rm GG}_{12}
					+ N^{\rm FF}_{12}w^{\rm FF}_{12}
					+ N^{\rm GF}_{12}w^{\rm GF}_{12}
					+ N^{\rm GF}_{21}w^{\rm GF}_{21}.
\end{split}
\end{equation}

The excess heat capacity $C(T)$ is calculated from the variance of equilibrium fluctuations of the total lattice enthalpy $H$ as follows:
\begin{equation}
C(T) = {\left\langle \left(H -\langle H \rangle  \right)^2 \right\rangle}/{RT^2}.
\end{equation}

The lipid interaction parameters were adjusted using the approach previously described by Sug\'{a}r \textit{et al.} \cite{Sugar1999, Sugar1994}: temperature-dependent excess heat capacity curves were obtained from MC simulations for a range of membrane compositions and compared with experimentally measured heat capacity curves, and the parameters $w^{mn}_{ij}$ were varied until a reasonable agreement with experimental DSC data was achieved. Lipid interaction parameters for our model are listed in Table \ref{Table1}. Since a simpler lattice representation of the lipid system (lipid molecules arranged on a square lattice) was used in our simulations compared to the previous studies \cite{Sugar1999, Sugar2000, Hac2005, Sugar1994} (lipids represented as dimers of acyl chains arranged on a triangular lattice), not surprisingly, the lipid interaction parameters $w^{mn}_{ij}$ in our study differ from those used in previous publications.

As is evident from Fig. \ref{FigS1}, with this set of lipid interaction parameters, the heat capacity-based empirical phase diagram obtained in our MC simulations agrees well with experimental data (both ours and previously published), as well as with previously published results of lattice-based MC simulations of the same lipid system.

\begin{table}[!t]
\caption{Thermodynamic parameters of lattice-based MC simulations of the DMPC/DSPC lipid membranes used in the present work.}
\label{Table1}
\begin{tabular*}{\textwidth}{@{}c@{\extracolsep\fill}c@{\extracolsep\fill}c@{\extracolsep\fill}c@{\extracolsep\fill}c@{\extracolsep\fill}c@{\extracolsep\fill}c@{\extracolsep\fill}c@{\extracolsep\fill}c@{\extracolsep\fill}c@{}}

\toprule

\multicolumn{10}{c}{Thermodynamic Parameters}\\

\midrule

 $\Delta E_1$
&$\Delta E_2$
&$\Delta S_1$
&$\Delta S_2$
&$w^{\rm GF}_{11}$
&$w^{\rm GF}_{22}$
&$w^{\rm GF}_{12}$
&$w^{\rm GF}_{21}$
&$w^{\rm GG}_{12}$
&$w^{\rm FF}_{12}$\\

 \tiny{(${\rm J}{\rm mol}^{\textnormal{-}1}$)}
&\tiny{(${\rm J}{\rm mol}^{\textnormal{-}1}$)}
&\tiny{(${\rm J}{\rm mol}^{\textnormal{-}1}{\rm K}^{\textnormal{-}1}$)}
&\tiny{(${\rm J}{\rm mol}^{\textnormal{-}1}{\rm K}^{\textnormal{-}1}$)}
&\tiny{(${\rm J}{\rm mol}^{\textnormal{-}1}$)}
&\tiny{(${\rm J}{\rm mol}^{\textnormal{-}1}$)}
&\tiny{(${\rm J}{\rm mol}^{\textnormal{-}1}$)}
&\tiny{(${\rm J}{\rm mol}^{\textnormal{-}1}$)}
&\tiny{(${\rm J}{\rm mol}^{\textnormal{-}1}$)}
&\tiny{(${\rm J}{\rm mol}^{\textnormal{-}1}$)}\\

\midrule

 26330
&50740
&88.653
&154.695
&1827
&1622
&4025
&4460
&1412
&502\\

\bottomrule
\end{tabular*}
\end{table}

For every lipid composition and temperature studied, the membrane was first equilibrated, if required. To do that, a system being initially at $T = \infty$ was brought to the equilibrium at a target temperature $T$ by exercising the Monte Carlo procedure which additionally included a third MC substep consisting in position exchange of two randomly chosen lipid molecules. This approach, which was suggested in \cite{Sugar1997} and successfully applied in \cite{Sugar1999,Sugar2000,Sugar2005} is known to be very efficient in driving the system toward the equilibrium configuration. This procedure was propagated typically for $1.5 \times 10^5$ MC cycles, which is substantially longer than the typical time required by the total lattice enthalpy $H$ to reach its equilibrium value at a given temperature in the presence of the random lipid exchange substep. After the completion of this procedure, the random lipid exchange substep is switched off, and the equilibrated system is ready for studies of its equilibrium properties.

The simulation code was written in Fortran and compiled with Compaq Visual Fortran Compiler Version 6.6A (Compaq Computer Corporation, Houston, TX). Results were analyzed using original dedicated routines written in MATLAB (The MathWorks, Nattick, MA). A reliable random number generator is essential for the success of an MC simulation. Therefore, we used the Mersenne Twister routine \cite{Matsumoto} providing sequences of pseudo-random numbers equidistributed in 623 dimensions and characterized by an extremely long period of $2^{19937} - 1 \approx 4.3 \times 10^{6001}$. Monte Carlo simulations were carried out on a workstation (Intel Core2 Quad Extreme CPU X9770 3.2 GHz, 4 GB RAM) running under Windows XP. Under these conditions, a simulation run on a $600 \times 600$ square lattice for $2 \times 10^{7}$ MC cycles takes about 700 h of CPU time.

\subsection*{Calculation of static structure factors for lipids and lipid states}
To gain information on the character of spatial fluctuation of the local membrane composition and lipid state in the equilibrium, the static structure factors for lipids (DMPC and DSPC) $S_\text{L}(k)$ and lipid states (fluid and gel) $S_\text{S}(k)$ were calculated independently.

The structure factor
\begin{equation}
S(k) = 1 + \bar{\rho}\hat{g}(k)
\end{equation}
is expressed via the Fourier transform $\hat{g}(k)$ of the pair distribution function of particles 
\begin{equation}
g(r) = \bar{\rho}^{-2}\left\langle \sum_i \sum_{j\neq i} \delta({\bf r}_i)\delta({\bf r}_j-{\bf r})\right\rangle ,
\end{equation}
where $\bar{\rho}$ is the spatially averaged number density of particles \cite{Allen}. To avoid artifacts in spatial and angular averaging, the fact that the particles cannot take arbitrary positions, but rather occupy sites on a square lattice, was taken into account.

\end{document}